# An emotional expression system with vibrotactile feedback during the robot's speech

Yuki Konishi[1], Yoshihiro Tanaka[1,2]

[1] *Nagoya Institute of Technology, Gokiso-cho, Showa-ku, Nagoya, Japan*

[2] *Inamori Research Institute for Science, Japan*

(Email: tanaka.yoshihiro@nitech.jp)

**Abstract** --- This study aimed to develop a system that provides vibrotactile feedback corresponding to the emotional content of text when a communication robot speaks. We used OpenAI's "GPT-4o Mini" for emotion estimation, extracting valence and arousal values from the text. The amplitude and frequency of vibrotactile stimulation using sine waves were controlled on the basis of estimated emotional values. We assembled a palm-sized tactile display to present these vibrotactile stimuli. In the experiment, participants listened to the robot's speech while holding the device and then evaluated their psychological state. The results suggested that the communication accompanied by the vibrotactile feedback could influence psychological states and intimacy levels.

## 1 Introduction

In recent years, advancements in robotics and AI have enabled communication robots to be active in various settings. While visual and auditory cues are the primary means of human communication, tactile sensations may also be related to emotions and interpersonal relationships [1][2]. Similar to human-to-human interactions, robot-human communication involving tactile interaction has also been investigated [3][4]. Many previous studies have focused on equipping robots with tactile sensors to evaluate tactile input and haptic interaction through active touch whereas presenting tactile stimulation is also promising to improve human-robot communication as it is available to any body area by using tactile displays. Thus, in this study, we developed a tactile communication system that provides vibrotactile stimuli corresponding to the robot's speech, modulating the stimuli on the basis of the emotion estimated from the text of the speech.

## 2 Proposed system

### 2.1 System configuration

This study aimed to develop a communication system in which vibrotactile stimulation is provided simultaneously with the speech of a communication robot. Figure 1 shows the configuration and appearance of the proposed system. In this system, vibrotactile stimulation is generated in real-time according to the emotion of the input text and presented with a tactile display along with the speech. The communication robot used was a humanoid robot Sota (Vstone) In the present system, the robot's speech was conducted in Japanese.

### 2.2 Emotion Estimation of Text

The input text was analyzed to estimate its emotional content, and the resulting emotional values determined the vibrotactile stimulation patterns presented. For emotion estimation, we used OpenAI's "GPT-4o Mini" via an API. The prompt was provided in Japanese to input the objective, teaching an explanation of Russell's circumplex model, procedures and criteria, output format, and important considerations [5]. The objective stated that the goal was to quantitatively assess the emotion of the input text based on Russell's circumplex model and output the emotional values.

The explanation of Russell's circumplex model indicated that it is a psychological model that evaluates emotions along two dimensions: valence: pleasant-unpleasant and arousal: activation-deactivation. Additionally, it was explained that each axis intersects at right angles, and emotions are arranged in a circular manner around the intersection point, with labels such as pleasure (0°), excitement (45°), arousal (90°), distress (135°), misery (180°), depression (225°), sleepiness

(270°), and contentment (315°). In the procedures and criteria section, we instructed the model to assess the valence dimension based on the frequency of positive or negative words and the overall tone of the text, and the arousal dimension based on how lively or calm the content was. The output format section instructed the model to output the emotional values in the format "Pleasure: XX.X%, Misery: YY.Y%, Arousal: ZZ.Z%, Sleepiness: AA.A%", ensuring that the total percentage of each axis sums to 100%. The important considerations section emphasized adhering to the output format and the explanation of Russell's circumplex model, and noted that the input text would be in Japanese. The above prompt allowed us to output the pleasure-misery and arousal-sleepiness values from the input text.

**2.3 VIBROTACTILE STIMULATION**

Vibrotactile stimulation was designed with a sinusoidal wave. The valence value, ranging from 100 (pleasant) to 0 (unplesant), was linearly mapped to a frequency range of 500-60 Hz as follows:

$$Freq = 60 + V(500 - 60)/100, \quad (1)$$

where Freq and V indicate the frequency and valance value, respectively. Similarly, the amplitude of vibrotactile stimulation, represented as 16-bit integer value in response to the arousal value increasing from 0 (deactivation) to 100 (activation), corresponding to a range from 8000 to 32767 as follows:

$$Amp = 8000 + A(32767 - 8000)/100, \quad (2)$$

where Amp and A indicate the amplitude set to PC and arousal value, respectively.

The vibrotactile stimulation data was generated as audio data using the Python library PyAudio and was played back using the tactile display. The presentation time was set to a basic duration of 0.5 seconds, with an additional 0.1 seconds per hiragana character and 0.2 seconds per kanji character, adjusting according to the length of the text. The vibrotactile feedback began simultaneously with the start of Sota's speech.

**2.4 TACTILE DISPLAY**

The tactile display was spherical as shown in Fig. 1, with a diameter of 80 mm, covered in black felt on the surface, and contained a vibrotactile actuator (foster 639897) inside.

**3 EXPERIMENTAL METHODS**

The experiment was conducted with four male and two female participants in their twenties. During the experiment, participants held the tactile display with both hands and listened to Sota's speech under two conditions: with vibrotactile feedback and without. The following ten short phrases were used in random order for each condition

1. "I'm happy you're listening to me."
2. "Ouch, ouch! Don't hit me."
3. "I had a bit of a scary dream."
4. "I'm hungry, but I can't eat because I'm a robot."
5. "It was a calm day today."
6. "I have nothing to do, and I'm feeling sleepy."
7. "I feel a bit foggy, like my mind isn't clear."
8. "Nothing special happened today."
9. "Every day is just so much fun!"
10. "I feel like I'm under a lot of stress. "

Half of the participants first listened to the ten phrases under the condition with vibrotactile feedback, while the other half experienced the reverse order. After listening to each phrase, participants were asked to rate their emotions using the Self-Assessment Manikin (SAM) [6]. At the end of each condition, they evaluated their psychological distance using the Inclusion of Other in the Self (IOS) scale [7]. Additionally, participants were asked to provide free-form comments at the end of the experiment.

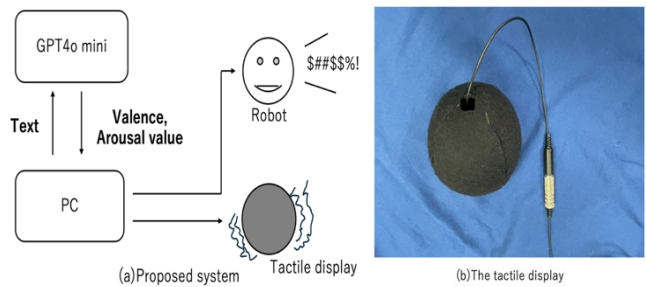

Fig.1 System configuration

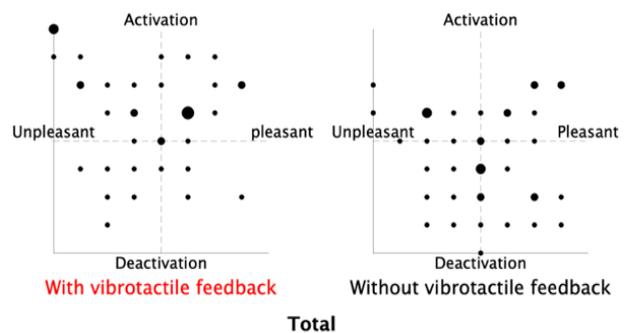

Fig.2 All result of the SAM test

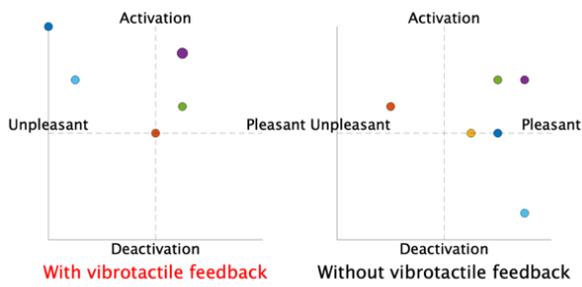
1. I'm happy you're listening to me.

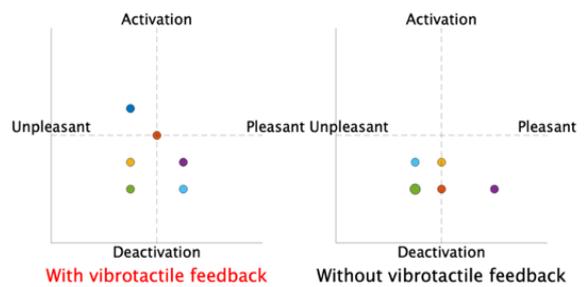
6. I have nothing to do, and I'm feeling sleepy.

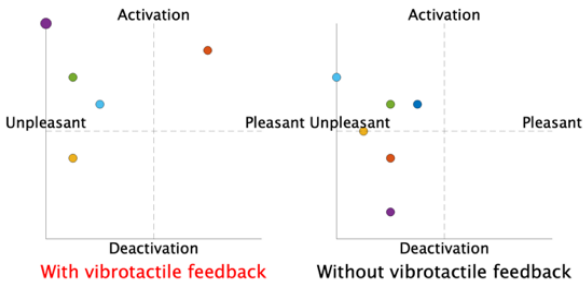
2. Ouch, ouch! Don't hit me.

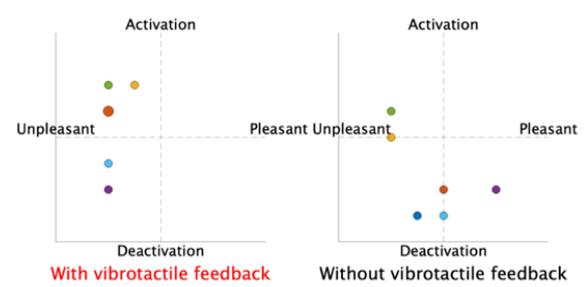
7. I feel a bit foggy, like my mind isn't clear.

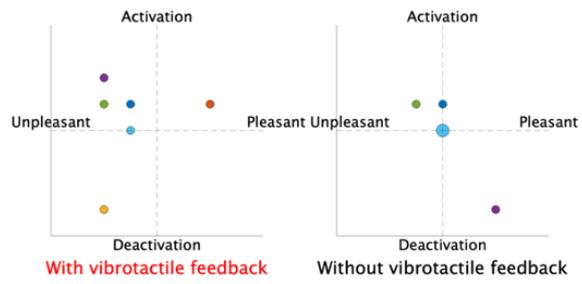
3. I had a bit of a scary dream.

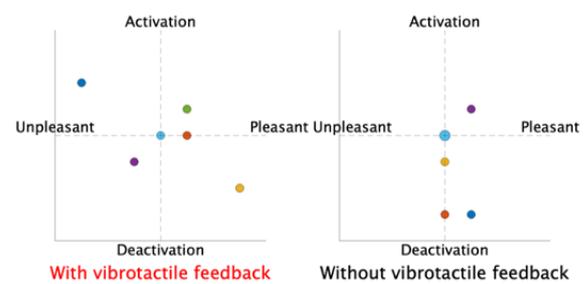
8. Nothing special happened today.

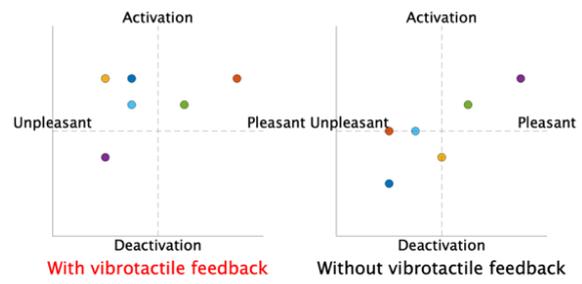
4. I'm hungry, but I can't eat because I'm a robot.

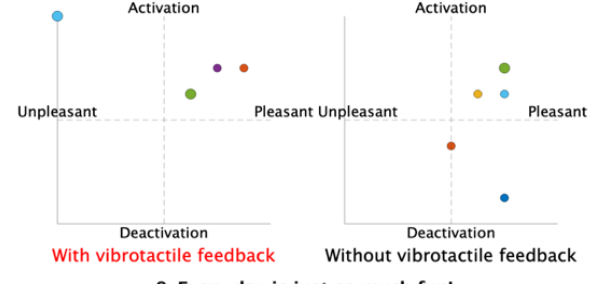
9. Every day is just so much fun!

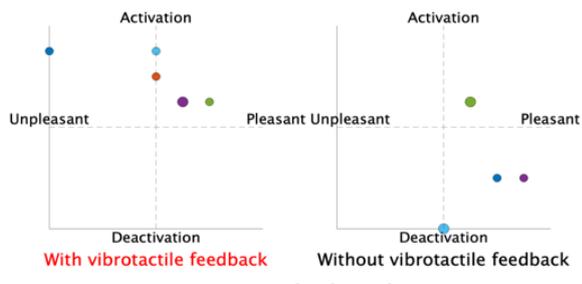
5. It was a calm day today.

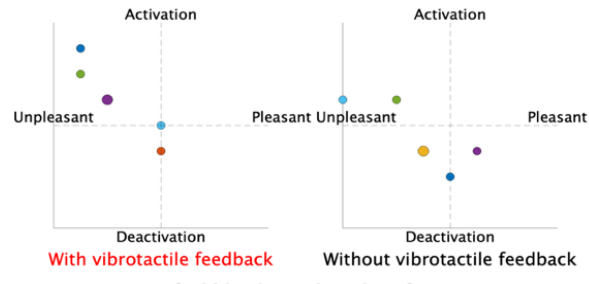
10. I feel like I'm under a lot of stress.

Fig.3 Result of SAM test for each phrase.
Different colors indicate each participant.

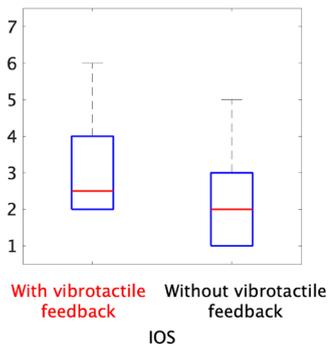

Fig.4 IOS

## 4 Results and discussion

Figure 2 shows all results of the SAM test involving all phrases. Figure 3 shows the results of the SAM test under the conditions with and without vibrotactile feedback for each phrase. The plot colors denote each participant and the diameter indicates the number of responses from participants. According to Figures 2 and 3, it can be observed that the valence score trended to be more unpleasant in the condition with vibrotactile feedback. Specifically, in phrases with negative emotions, such as phrases No. 2, 3, and 8, unpleasant ratings increased with vibrotactile feedback. Additionally, even in phrases with positive emotions, such as phrases No. 1 and 9, unpleasant ratings could increase with vibrotactile feedback. Overall, arousal rantings were higher with vibrotactile feedback than without that.

Figure 4 shows the results of the IOS test. It was found that the intimacy scores were tended to be higher in the condition with vibrotactile feedback.

In addition to these psychological tests, interesting trends emerged from the participants' free-form comments. Two participants noted that "the combination of statements like 'Ouch, ouch!' or ' I had a bit of a scary dream' with low-frequency vibrations evoked strong or unpleasant emotions." Another two participants mentioned that "the high-frequency vibrations felt uncomfortable despite the positive nature of the statements." These comments support the results showing unpleasant ratings in the condition with vibrotactile feedback. Other comments included "I was surprised at how much the vibrations changed my feelings of pleasure/displeasure and arousal" and "the vibrations made me more aware of Sota." These responses support the observed changes in arousal levels. Regarding intimacy, one participant mentioned "Sota is cute," which corresponds to the maximum IOS score among participants in both conditions. However, there was also a comment from one participant stating, "The vibration interfered with my sense of intimacy with Sota," which contradicted the overall trend, implying individual differences in interpretation and influence of vibrotactile feedback.

## 5 Conclusion

This study developed a communication system in which vibrotactile stimulation is provided simultaneously with the speech of a communication robot. From the results of the psychological tests and the free-form comments, it is suggested that vibrotactile feedback during the robot's speech can potentially influence the psychological state of participants. In future work, we will improve the present system by investigating different types of vibrotactile stimulation. Additionally, long-term studies will be investigated to understand how the system affect over time in daily life.

## Acknowledgement

This work was supported in part by Inamori Research Institute for Science, JSPS Grant-in-Aid for Scientific Research (JP24H00741), and the commissioned research by National Institute of Information and Communications Technology, Japan.